\documentclass[journal=jpccck,manuscript=article]{achemso}
\usepackage{achemso}
\usepackage{amsmath}

\author{Meijuan Cheng}
\affiliation
{Department of Physics, Key Laboratory of Low Dimensional Condensed Matter Physics
, Xiamen University, Xiamen 361005, China}
\author{Xiaohong Shi}
\affiliation
{Department of Physics, Key Laboratory of Low Dimensional Condensed Matter Physics
, Xiamen University, Xiamen 361005, China}
\author{Shunqing Wu}
\affiliation
{Department of Physics, Key Laboratory of Low Dimensional Condensed Matter Physics
, Xiamen University, Xiamen 361005, China}
\author{Zi-Zhong Zhu}
\affiliation
{Department of Physics, Key Laboratory of Low Dimensional Condensed Matter Physics
, Xiamen University, Xiamen 361005, China}
\affiliation
{Fujian Provincial Key Laboratory of Theoretical and Computational Chemistry,
Xiamen 361005, China}
\email{zzhu@xmu.edu.cn}

\title[An \textsf{achemso} demo]
  {Significant Second-Harmonic Generation and Bulk Photovoltaic Effect in Trigonal Selenium and Tellurium Chains }

\abbreviations{IR,NMR,UV}

\begin{document}

%
%
%
%
%

\begin{abstract}
One-dimensional selenium and tellurium crystalize in helical chainlike structures and thus exhibit fascinating properties. By performing first-principles calculations, we have researched the linear and nonlinear optical (NLO) properties of 1D Se and Te, and find that both systems exhibit pronounced NLO responses. In particular, 1D Se is found to possess large second-harmonic generation coefficient
with the $\chi^{(2)}_{xyz}$ being up to 7 times larger than that of GaN, and even is several times larger than that of the bulk counterpart. On the other hand, 1D Te also produces significant NLO susceptibility $\chi^{(2)}_{xyz}$ which exceeds the bulk GaN by 5 times. Furthermore, 1D Te is shown to possess prominent linear electro-optic coefficient $r_{xxx}(0)$. Particularly, Te chain exhibits large shift current response and the maximum is more than the maximal photovoltaic responses obtained from BaTiO$_3$ by 2 times. Therefore, 1D Se and Te may find potential applications in solar energy conversion, electro-optical switches, and so on. Finally, much stronger NLO effects of 1D Se and Te are attributed to their one-dimensional structures with high anisotropy, strong covalent bonding and lone-pair electrons. These findings will contribute to the further study in experiments and search for excellent materials with large NLO effects.
\end{abstract}
\newpage
\section{INTRODUCTION}
In the strong optical fields, materials with broken inversion symmetry will generate large nonlinear optical effects \cite{Shen2003,Boyd2003} such as second-harmonic generation (SHG), linear electro-optic (LEO) effect, bulk photovoltaic effect (BPVE). Second-harmonic generation(SHG), the best-known nonlinear optical (NLO) responses, have promising applications in probes of surfaces and interfaces, frequency doublers, and so on. Notably, it has been investigated extensively in bulk semiconductors \cite{Boyd2003,Chang1965,zhong1993,Hughes1996,Gavrilenko2000,Cai2009,Cheng2019,Ni2020} for nearly
five decades beginning in the early 1960s. More recently, the family of NLO materials has also been extended to one-dimensional (see, e.g., refs 10-11 and references therein) and two-dimensional (see, e.g., refs 12-17 and references therein) systems, crucial for the development modern optical and electro-optical devices such as frequency conversions, electro-optic modulators, and switches \cite{Boyd2003}. As
another second-order electric polarization effect, linear electro optic (LEO) effect refers to the linear refractive index variation ($\Delta n$) with the applied electric field strength ($E$), i.e., $\Delta n = n^3rE/2$, here $r$ is the LEO coefficient and $n$ is the refraction index \cite{Boyd2003}. Therefore,
the LEO effect could allow one to use an electrical signal to control the amplitude, phase, or direction of a light beam in the NLO materials, and leads to a widely used means for high-speed optical modulation and sensing devices (see, e.g., ref 18 and references therein). And more notably, the bulk photovoltaic
effect (BPVE), a third nonlinear optical response, has become a vibrant research topic in recent years. Unlike traditional photovoltaics, the BPVE can be obtained in single crystal with the broken inversion symmetry, which is extremely ponderable for solar energy collection and conversion.

The crystal structures of trigonal selenium and tellurium consist of the helical chains arranged in a hexagonal array, with their axes parallel to the c axis. Along the chain, every atom is covalently bonded to two neighboring atoms, and interacts with four second nearest-neighbor atoms of the adjacent chains by relatively weaker interaction, or more precisely, covalent-like quasi-bonds (CLQB) \cite{qiao2018}, thus indicating that they may allow to be exfoliated as the individual atomic chains. Excitingly, Li {\it et al.} reported the fabrication process of single Se chains and the experimental study of the geometry, phase stability, electronic properties, and so on \cite{Li2005}. More recently, Churchill {\it et al.} demonstrated the potential for fabrication of isolated tellurium atom chains by mechanical exfoliation in experiment \cite{Churchill2017}. The experimental breakthrough triggered a growing renewed interest in 1D selenium and tellurium \cite{Olechna1965,Springborg1988,Ghosh2007,Kahaly2008,Tuttle2017,Andharia2018,Pan2018}.
In particular, Se and Te chains maintain the interesting nature of Se and Te bulks, i.e., broken spatial inversion symmetry, and thus make the 1D systems possess fascinating properties such as nonlinear optical responses, piezoelectric and ferroelectric properties.

As elemental semiconductor materials, 1D selenium and tellurium chains has been investigated extensively in the structural, electronic, and linear optical properties. \cite{Olechna1965,Springborg1988,Ghosh2007,Kahaly2008,Tuttle2017,Andharia2018,Pan2018} However, there is no systematic study on second-order nonlinear optical responses of these helical atomic chains to date. It is well known that low dimensional semiconductors could possess attractive properties not seen in their bulk counterparts, for example, spin-valley coupling \cite{Xiao2012}, and further have attracted widespread attention in recent years. Indeed, having band gap in the visible spectrum (1.6-3.1 eV), the Se and Te chains could be expected to show significant BPVE response and then equip potential applications in solar energy harvesting.

In this work, we perform first principles calculations of the linear dielectric functions, the second-order NLO susceptibility, linear electro-optic effect and the bulk photovoltaic effect for 1D selenium and tellurium. Our main purpose is to find out the magnitude and characteristics of NLO effects in both systems and thus predict whether they have any potential applications in the domain of nonlinear optics such as electro-optic modulator, frequency multiplier, optical switching, and sum-frequency generation. Accordingly, to motivate the potential applications and experimental developments of Se and Te chains in the near future, we systematacially investigate the NLO properties of these chain structures.

\section{COMPUTATIONAL DETAIL}
In this paper, we study the electronic property, optical dielectric function, SHG and LEO coefficients over the entire optical frequency range, as well as the shift current (SHC) response. The crystal structure of 1D selenium or tellurium is schematically displayed in Fig. 1. It consists of three atoms in the unit cell situating at positions ($u, 0, 0$), ($0, u, 1/3$) and ($-u, -u, 2/3$). The valence electron configurations of 1D selenium and tellurium are $4s^2$$4p^4$ and $5s^2$$5p^4$, respectively. That is one third of the p bands are empty and thus every atom is covalently bonded to two neighboring atoms along helical chain. For selenium and tellurium bulks, coordination number of atoms is 6, whereas that of single helical chain reduces to 2. Consequently, intrachain distances in single helical chain are shorter than that of bulks due to the lower coordination number of atoms (see Table 1). The 1D structures are modeled by using the slab-superlattice approach with a 20 \AA\  thickness of the vacuum region along the a and b-direction, ensuring negligible interaction between the periodic images. The effective unit cell volumes of 1D systems are expressed as $V_{eff}$ = ($V$$\cdot$$a$$\cdot$$b$)/(20\AA$\cdot$20\AA), here $a$ and $b$ are corresponding lattice constants of bulks, rather than the arbitrary volume of the supercells $V$.

\begin{table}
\caption{Calculated lattice constant $c$, the distances of intrachain $r$, effective unit cell volume $V_{eff}$, calculated band gap with GGA ($E_{g}^{GGA}$)
and the HSE calculations with the SOC ($E_{g}^{HSE-SOC}$) as well as scissors operator ($\Delta E_g = E_{g}^{HSE-SOC} - E_{g}^{GGA}$) for  selenium and tellurium chains.
}
\begin{tabular}{c c c c c c c}
         & $c$ (\AA) & $r$ (\AA) & $V_{eff}$ (\AA$^3$) & $E_{g}^{GGA}$ (eV) & $E_{g}^{HSE-SOC}$ (eV) &   $\Delta E_g$ (eV)  \\  \hline
 Se bulk &4.954\textsuperscript{\emph{a}}& 2.373\textsuperscript{\emph{a}}& 81.78\textsuperscript{\emph{a}}&1.002\textsuperscript{\emph{a}}&1.735\textsuperscript{\emph{a}}&0.733\textsuperscript{\emph{a}}   \\
 Se chain&4.956 (4.949)\textsuperscript{\emph{c}}&2.363& 81.82&1.992(2.043)\textsuperscript{\emph{c}}  &2.859   &   0.867    \\
 Te bulk &5.926\textsuperscript{\emph{a}}& 2.833\textsuperscript{\emph{a}}&101.68\textsuperscript{\emph{a}}&0.113\textsuperscript{\emph{a}}&0.322\textsuperscript{\emph{a}}&0.209\textsuperscript{\emph{a}}   \\
 Te chain&5.690(5.69)\textsuperscript{\emph{b}}(5.651)\textsuperscript{\emph{c}} & 2.747 (2.74)\textsuperscript{\emph{b}} & 97.63& 1.666(1.750)\textsuperscript{\emph{c}} &2.194 & 0.528  \\
\hline
\end{tabular}

\textsuperscript{\emph{a}} Reference 8.

\textsuperscript{\emph{b}} Reference 24.

\textsuperscript{\emph{c}} Reference 27.
\end{table}

Our ab initio calculations are performed using the accurate projector augmented wave (PAW) method \cite{blochl1994}, as implemented in the Vienna ab initio simulation
package\cite{kresse1996ab,kresse1996}, They are based on the density-functional theory with the generalized gradient approximation (GGA) of Perdew, Burke, and Ernzerhof \cite{perdew1996}.
An adequately large plane-wave cutoff ($E_{cut}$) of 450 eV is adopted throughout and the total energy convergence criterion for the self-consistent electronic
structure calculations is $10^{-6}$ eV. The atomic positions and lattice constants are fully relaxed until the forces acting on all the atoms are less than
0.001 eV/\AA. $K$-point meshes of 1$\times$1$\times$40 for Se and Te single helical chains are used for the self-consistent charge density calculations.
Further calculations using different k-point meshes indicate that the above k-point meshes yields the well-converged charge density.

In our work, the linear optical dielectric function, second-harmonic generation (SHG), linear electro-optic (LEO) effect and the shift current (SHC) response
are calculated based on the linear response formalism with the independent-particle approximation, as described previously \cite{guo2004,guo2005ab}.
The imaginary part of the dielectric function $\varepsilon(\omega)$ is given by the Fermi golden rule due to direct interband transitions.
The real part of the dielectric function is obtained from the calculated $\varepsilon''(\omega)$ by the Kramer-Kronig transformation. As previously reported, the imaginary part of the second-order optical susceptibility due to direct interband transitions is given by \cite{guo2004,guo2005}
\begin{equation}
\chi''^{(2)}_{abc}(-2\omega,\omega,\omega) = \chi''^{(2)}_{abc,VE}(-2\omega,\omega,\omega)+\chi''^{(2)}_{abc,VH}(-2\omega,\omega,\omega),
\end{equation}
where the contributions are derived from the so-called virtual-electron process $\chi''^{(2)}_{abc,VE}$ and the virtual-hole process $\chi''^{(2)}_{abc,VH}$. \cite{guo2004,guo2005}
Then the real part of the second-order optical susceptibility $\chi_{abc}'^{(2)}$ is obtained from the calculated imaginary part $\chi''^{(2)}_{abc}$ by the Kramer-Kronig transformation.
The linear electro-optic $r_{abc}(\omega)$ is related to the second-order optical
susceptibility $\chi_{abc}^{(2)}(-\omega,\omega,0)$\cite{Hughes1996}. In the zero frequency limit, $r_{abc}(0)$  is expressed as
\begin{equation}
r_{abc}(0)=-\frac{2}{\varepsilon_a(0)\varepsilon_b(0)}\lim_{\omega\rightarrow 0}\chi_{abc}^{(2)}(-2\omega,\omega,\omega).
\end{equation}
For the photon energy $\omega$ well below the band gap, $\chi_{abc}^{(2)}(-2\omega,\omega,\omega)$  and $n(\omega)$ are nearly constant
and then the LEO coefficient $r_{abc}(\omega)\approx r_{abc}(0)$\cite{wang2015}. Furthermore, we calculate the major origins of the bulk
photovoltaic effect, i.e., the shift current, by the following equation \cite{Sipe2000},
\begin{equation}
J_{a}(\omega)=\sum_{bc}\sigma_{abc}(\omega)E_b(\omega)E_c(\omega).
\end{equation}
Where the third-rank tensor $\sigma_{abc}(\omega)$ could be given by using the expression from ref 32.

In order to obtain accurate optical properties, several different $k$-point grids are calculated for selenium and tellurium chains until
the calculated optical properties converge to a few percent. In conclusion, we use a denser enough k-point meshes of 2$\times$2$\times$120
for 1D Se and Te chains. Test calculations indicate that using 30 bands per atom included in the optical calculations is sufficiently
accurate for $\varepsilon'(\omega)$ and $\chi_{abc}'^{(2)}$ obtained by the Kramer-Kronig transformation. Furthermore, the calculated
optical spectra are broadened by using a Gaussian function with $\Gamma=0.2$ eV.

As mentioned above, our calculated linear and nonlinear optical properties are based on the independent-particle approximation (IPA),
i.e., the quasi-particle self-energy corrections and excitonic effects were neglected. However, low dimensional materials possess quantum
confinement and thus many-body effects play a remarkable role on the linear optical properties of low dimensional systems such as MoS$_2$ ML, SiC sheet
\cite{Hsueh2011,Qiu2013}. It is widely known that precise first-principles calculations considering the excitonic effect such as GW approximation
combined with the  Bethe-Salpeter equation (BSE) could produce sufficiently accurate optical spectra compared to experimental measurements. \cite{Qiu2013}
Nevertheless, these approaches are performed on the basis of several thousand $k$-points and tens of bands and then are usually extremely demanding
computationally. Worse still, the expression for nonlinear susceptibility is much more complicated than that of the linear optic, resulting in
failure to perform the full first-principles calculations with many-body effects included. As a result, much simpler approaches such as the
real-time propagation, \cite{Gru14} scissors correction (SCI) \cite{wang2015} and semiempirical tight-binding-based model potential \cite{ Trolle2014}
were used to calculate nonlinear optical property. Similarly, we take the self-energy corrections into account by the so-called scissors correction
(SC) to reduce the errors of calculated linear dielectric function and NLO coefficients caused by neglected many-body effects. Indeed, such SC
calculations have been testified to generate the NLO susceptibility at zero frequency for low-dimensional
materials such as graphene-like BN sheets \cite{wang2015} that agree well with the experimental ones.

\section{RESULTS AND DISCUSSION}

\subsection{Band Structures of Se and Te Chains}

The calculated band structures of selenium and tellurium chains are extremely similar due to the same crystalline structures, as displayed in Figure 2.
Both systems are indirect band-gap semiconductors with the conduction band minimum (CBM) being at the A point, whereas the valence band maximum (VBM)
is located somewhere slightly closed to A along the $\Gamma$-A symmetry line. The valence bands of two materials could be divided into three groups, arising
from s bonding, p bonding, and p lone pair states, respectively, and the three lowest conduction bands are corresponding to the p antibonding states.
Compared to bulk systems, less electrons transferred from lone pair states to antibonding states since interchain interaction is not reside in
one-dimensional chain structures. This explains why the calculated band gap of 1D Se and Te is much larger than that of bulk counterparts, especially
for tellurium chain. Obviously, the band gap we calculated with GGA agrees rather well with previous results \cite{Andharia2018} (see Table 1).

As mentioned previously \cite{Cheng2019}, the band gaps calculated by the GGA are generally underestimated. Thus, to obtain accurate optical properties
that are on the basis of precise band structure, we also calculate the band structures using the hybrid Heyd-Scuseria-Ernzerhof (HSE) functional \cite{heyd2003}
with the SOC included, as shown in Fig. 2. Indeed, in our precious report \cite{Cheng2019}, the comparison of HSE-SOC band gaps and experimental values
for bulks proves that the HSE calculations with the SOC is sufficiently accurate in describing band gap. Therefore, all the optical properties in 1D
materials are calculated from the GGA band structures based on the scissors correction (SCI) \cite{wang2015}. The scissors correction energy, i.e., the
differences between the HSE-SOC and GGA band gaps, and HSE-SOC band gaps are listed in Table 1.

\subsection{Linear Optical Property of Se and Te Chains}

The calculated imaginary and real part of dielectric function for one-dimensional selenium and tellurium are displayed in Fig. 3.
Similar to bulk systems, 1D counterparts are uniaxial crystalline structures with strongly covalent bonded helical chains oriented
along the c axis and then their optical properties are strongly dependent on light polarization direction.  As a result, the constituent
elements of dielectric function are two significantly different components, namely light polarization parallel ($E \parallel c$) and
perpendicular ($E \parallel a$) to the c axis. Apparently, the spectral feature of dielectric function for selenium chain is extremely
similar to that of tellurium due to the same crystalline structures. Hence, in what follows, we will take that of 1D Se as an example
to perform a detailed analysis [see Fig. 3(a) and (b)]. As expected, stronger anisotropy in the spectra of dielectric function can be
observed. This could be explained by the fact that the two systems possess a much stronger structural anisotropy than bulks. Figure 3(a)
shows that the spectra of the imaginary part of the dielectric function for Se chain could be conveniently divided into two regions.
Due to different optical selection rules, in the low-energy region (about 2.5-5.5 eV), there is prominent optical absorption for $E \parallel c$,
whereas, the greatly weak absorption is observed in the range of energy for $E \parallel a$. Furthermore, our calculations predict two rather
remarkable peaks in the imaginary part of the dielectric function for $E \parallel c$ (direction of chains) within the energy region, i.e., a much
larger peak located at $\sim$3.9 eV and a slightly small one in the vicinity of 3.1 eV. In the energy range from 5.5 to 14.0 eV, one-dimensional
selenium exhibits a pronounced peak at $\sim$7.5 eV and a relatively small peak near 6.8 eV in the spectrum of $\varepsilon''(\omega)$ for $E \parallel c$. However,
the spectrum of $E \parallel a$ exhibits some steadily oscillatory bulges and a relatively weak peak centered at $\sim$8.3 eV. In particular,
a deep minimum occurs at $\sim$5.5 eV originating from the fact that transitions from the upper valence triplet to the lower conduction triplet
are already exhausted. Beyond that, the increase in the spectral amplitude within the energy range between 5.5 eV and 14.0 eV is due to
the transitions from the lower valence triplet to the lower conduction triplet and also from the upper valence triplet to the upper conduction triplet.

\subsection{Second-Harmonic Generation and Linear Electro-Optic Coefficient of Se and Te Chains}

Similarly to bulk materials, one-dimensional selenium and tellurium also possess five nonzero nonlinear susceptibility and only two elements
are independent, i.e., $\chi^{(2)}_{xxx}=-\chi^{(2)}_{xyy}=-\chi^{(2)}_{yxy}$, $\chi^{(2)}_{xyz}=-\chi^{(2)}_{yzx}$. Notably, based on the
scissors correction, the peak positions are blue-shifted by about the energy of
the scissors correction ($\Delta E_g$) and the magnitude of the SH susceptibility gets reduced (see Table 2), while the line shapes of SHG
spectra are hardly changed. As a result, only the results obtained with scissors correction are displayed in this paper. The static values
of the dielectric constant, LEO coefficient and second-order NLO susceptibility in the 1D selenium and tellurium are listed in Table 2.
Compared with bulk Se and Te \cite{Cheng2019}, the static SHG coefficients for Se and Te chains decrease which, however, are comparable
to that of GaN in both zinc-blende and wurtzite structures \cite{Gavrilenko2000,Cai2009}. In addition, Te chain possesses large LEO
coefficient $r_{xxx}(0)$ ($\sim$1.5 pm/V), exceeding 2 times larger than that of bulk GaN polytypes \cite{Gavrilenko2000,Cai2009}.

\begin{table}
\caption{Static dielectric constants ($\varepsilon_x = \varepsilon_y$ and $\varepsilon_z$),
second-order susceptibility $\chi^{(2)}_{xxx}(0)$ (pm/V) and $\chi^{(2)}_{xyz}(0)$ (pm/V)
 as well as  LEO coefficient $r_{xxx}(0)$ (pm/V)
and $r_{xyz}(0)$ (pm/V) of selenium and tellurium chains calculated without (GGA) and with (SC) scissors correction.
}
\begin{tabular}{ c c c c c c c c }
  &  & $\varepsilon_x$ & $\varepsilon_z$ & $\chi^{(2)}_{xxx}(0)$  & $\chi^{(2)}_{xyz}(0)$ & $r_{xxx}(0)$ & $r_{xyz}(0)$   \\ \hline
 Se chain& GGA & 5.1 & 12.7 & 19  & 7   & -1.48  & -0.50    \\
         & SC & 4.6  & 10.3 & 9   & 2   & -0.87    & -0.16     \\
 Te chain& GGA & 6.8 & 16.7 & 54  & 49   & -2.33  & -2.10    \\
         & SC & 6.3  & 14.2 & 29   & 23   & -1.46    & -1.15    \\
\hline
\end{tabular}
\end{table}

The calculated real and imaginary parts as well as the absolutes values of the second-order susceptibility are presented in Figs. 4 and 5
for selenium and tellurium chains, respectively. Apparently, the real and imaginary parts of second-order NLO susceptibility for both systems
show an oscillatory behavior. Indeed, for Se chain, the spectrum of absolute value of the SH generation coefficients oscillates rapidly and
dominating values are distribute in the photon energy range of 1.0-7.0 eV [see Figs. 4(b) and 4(e)]. Furthermore, the absolute value of $\chi^{(2)}_{xyz}$
reaches the maximum $\sim$1502 pm/V at $\sim$3.0 eV which, notably, is nearly 7 times larger than that of GaN  \cite{Gavrilenko2000,Cai2009},
a widely used NLO semiconductor, and even is several times larger than that of bulk counterpart. This indicates that one-dimensional selenium
would be excellent nonlinear optical materials and have potential applications in the field of nonlinear optics such as second-harmonic
generation, frequency conversion, optical switching, optical modulation, and so on. Unlike Se chain, the oscillatory spectra of the
second-order NLO susceptibility in 1D Te are none the less significant over the entire optical frequency range and the maximal NLO
susceptibilities tensor $|\chi^{(2)}_{xyz}|$ is as high as 1219 pm/V at 2.34 eV, which exceeds 5 times larger than that of GaN \cite{Gavrilenko2000,Cai2009}.
Significantly, for both materials, the real part, imaginary part, and absolute value of $\chi^{(2)}_{xyz}$ are larger than $\chi^{(2)}_{xxx}$,
which is contrary to bulk counterparts. The phenomenon could be explained by the fact that the two systems are one-dimensional spiral chains along the $c$ axis
and have a high degree of anisotropy.

The SHG involves not only single-photon ($\omega$) resonance but also double-photon (2$\omega$) resonance. Therefore, to further analyze the NLO
responses and understand the prominent features in the spectra, we also exhibit the dielectric functions $\varepsilon''(\omega)$ and $\varepsilon''(\omega/2)$
together in Figs. 4 and 5 for 1D selenium and tellurium, respectively. Because of similar features, in the following section, we only take the second-order
susceptibility element $\chi^{(2)}_{xyz}$ of 1D Te as an example to perform a brief analysis. Figs. 5(e) and 5(f) show that the threshold of the $\chi^{(2)}(-2\omega,\omega,\omega)$
spectra is corresponding to the absorption edge of $\varepsilon''(\omega/2)$, i.e., $\sim$ 1.0 eV ($\sim$ $\frac{1}{2}E_{g}$), while the absorption
edge of $\varepsilon''(\omega)$ is close to the band gap value $E_{g}$ ($\sim$2.0 eV). Therefore, the second-order NLO susceptibility spectra can
be divided into two parts. The first part from $\sim$1.0 to $\sim$2.0 eV is formed by double-photon resonances. The second part (above $\sim$2.0 eV)
is associated single-photon resonances with contribution from double-photon resonances [see Fig. 5(f)]. Apparently, these two types of
resonances cause the SHG spectra to violently oscillate and finally decrease gradually in the higher energy region.

\subsection{Bulk Photovoltaic Effect of Se and Te Chains}

The bulk photovoltaic effect (BPVE) is of the utmost importance to solar energy harvesting, which can be observed in single-phase homogeneous materials
without inversion symmetry. Moreover, the shift-current (SHC) is dominating mechanism of the bulk photovoltaic effect (BPVE). Therefore, in our work,
we have calculated the SHC coefficients for one-dimensional selenium and tellurium. Due to specific symmetry class, the calculated SHC coefficients
possess same nonzero elements with that of second-order NLO susceptibility, i.e., $\sigma_{xxx}$=$-\sigma_{xyy}$=$-\sigma_{yxy}$, $\sigma_{xyz}$=$-\sigma_{yzx}$.
Similar to second-order NLO susceptibility, the spectra of the shift-current in both systems exhibit significant oscillatory behavior and distribute
in a broad range of energy, as displayed in Fig. 6. Furthermore, our calculations for two helical structures also reveal that the maximal value of
the shift current responses $\sigma_{xxx}$ is larger than that of $\sigma_{xyz}$, and is opposite in sign. Obviously, the threshold of shift current
spectra and the absorption edge of the imaginary part of the dielectric function are approximately equal, in spite of small and flat. The maximal
element $\sigma_{xxx}$ in 1D Se exhibits a prominent peak near 10.1 eV, which is larger than the maximum response observed for BaTiO$_3$ \cite{Young2012},
an archetypical single-crystal with the bulk photovoltaic effect. Interestingly, our calculations predict that Te chain exhibits more pronounced shift
current response and the maximum in SHC coefficient $\sigma_{xxx}$ is more than the maximal photovoltaic responses obtained from BaTiO$_3$ by 2 times \cite{Young2012}.
Overall, our results indicate that one-dimensional selenium and tellurium, especially for tellurium, are outstanding broadband photovoltaic materials
and possess potential application prospects in solar energy conversion. Therefore, one could expect our calculations would boost further research
on the experiments and potential applications. Furthermore, such remarkable SHC response can be attributed to strong covalent bond in 1D Se and Te which,
in what follows, will be explained in detail.

It is widely known that the smaller the band gap is, the larger the magnitude of the imaginary part of the dielectric function, and NLO responses would be,
as explained in our previous work \cite{Cheng2019}, only due to the energy differences between the initial and final states of optical excitations in the denominators.
Particularly, the magnitude of the imaginary part of the second-order NLO susceptibility, especially in the low frequency region, would be approximately proportional
to the inverse of the fourth power of the band gap. However, compared to common semiconductors with semblable band gaps  \cite{Reshak2005,Kong2014,Hu2017S,Hu2017pe,Hu2017,Jiang2020,Ouahrani2011,Ma2017,Lin2019},
one-dimensional selenium and tellurium exhibit much larger second-order NLO responses. To further investigate the origins of the enhanced NLO effects of one-dimensional systems,
we calculate the deformation charge density, as depicted in Fig. 7. Clearly, considerable electrons accumulate in the neighborhood of the Se-Se (Te-Te) bond center
by exhausting the electrons around the atoms in the direction of the bond, and then lead to the strongly directional covalent bonds. Strong covalency is beneficial
to large spatial overlap between the wave functions of initial and final states resulting in large optical matrix elements, and thus give rise to large SHG values.
Furthermore, Fig. 7 also displays charge accumulate around each atom in the direction perpendicular to the chain, indicating the presence of lone-pair electrons.
Notably, lone-pair electrons are to the benefit of the generation of induced dipole oscillations by the optical electric fields, thus resulting in enhanced NLO
responses \cite{Jiang2014,Cammarata2014}. There is, in addition, one further point is high anisotropy for one-dimensional materials, which would lead to large joint DOS, would bring
large $\chi^{(2)}$ into being \cite{Ingers1988,Song2009}. As a result, much larger linear and nonlinear optical responses in one-dimensional selenium and tellurium attribute to
strong covalency, lone-pair electrons, high anisotropy. Similar to band gap, the volume, another crucial quantity in optical effects, is also reflected in the
denominators of the equations for imaginary part of the dielectric function and second-order NLO susceptibility, thus suggesting that the large volume is to the
disadvantage of pronounced optical responses. Being compared with bulk counterpart, 1D Se chain possesses higher anisotropy and comparable volume, in spite
of increased band gap. This explains that selenium chain has larger SHG coefficients than that of selenium bulk. Conversely, even though 1D Te exhibits higher
anisotropy and comparable volume, the greatly enhanced band gap results in much smaller NLO values. Therefore, in general, to search for excellent NLO
materials, one could focus on those semiconductors with smallest possible band gaps which are larger than the optical frequencies required by specific
NLO applications, strong covalency, lone-pair electrons, high anisotropy and small volume.

\section{CONCLUTIONS}
Summarizing, we have performed a systematic first-principles calculations of the linear and nonlinear optical properties of one-dimensional selenium and tellurium based on DFT within the GGA plus scissors
correction. Compared to the common semiconductors with similar band gaps, 1D Se and Te chains exhibit enhanced SHG, LEO as well as SHC responses. Also, due to their
structural anisotropy, the linear and nonlinear optical effects manifest stronger anisotropy than the corresponding bulk systems. Especially, one-dimensional
selenium exhibits remarkable SH generation coefficient with the $\chi^{(2)}_{xyz}$ being nearly 7 times larger than that of GaN, a widely used NLO semiconductor.
On the other hand, 1D Te is found to possess large second-order NLO susceptibilities $\chi^{(2)}_{xyz}$ being as large as 1219 pm/V, which is more than five times larger than that of GaN. Furthermore, Te chain exhibits prominent LEO coefficient which exceeds the bulk GaN polytypes by 2 times. It's worth noting that 1D Te is shown to have large shift current (SHC) response and the maximum is more than the maximal photovoltaic responses obtained from BaTiO$_3$ by 2 times. The maximal shift current element in 1D Se is also larger than BaTiO$_3$, an archetypical single-crystal with the bulk photovoltaic effect. Thus, 1D Se and Te may have application potentials in solar energy conversion, second-order nonlinear optical devices and linear electro-optic modulators, and so on. The prominent structures in the $\chi^{(2)}$ spectra of 1D Se and Te have been successfully associated with single-photon and double-photon resonances. Finally, enhanced NLO effects in 1D selenium and tellurium compared to the general semiconductors with similar band gaps ascribe to their one-dimensional structures with lone-pair electrons, high anisotropy,
strong covalency. This also indicates a strategy to search for excellent NLO materials with a specified band gap, i.e., starting with chainlike semiconductors with strong covalency and/or lone-pair electrons, high anisotropy. We expect that our work will stimulate further experiments and practical application on the SHG, LEO and SHC responses in these lower-dimensional materials.

\begin{acknowledgement}
M. J. Cheng thanks G.-Y. Guo of Department of Physics and Center for Theoretical Physics, National Taiwan University for his hospitality during her three months visit there.
The work is supported by the National Key R$\&$D Program of China (Grant No. 2016YFA0202601),
and the National Natural Science Foundation of China (No. 11574257).
\end{acknowledgement}

%
%


\newpage

\begin{figure}
\includegraphics[width=3.33in]{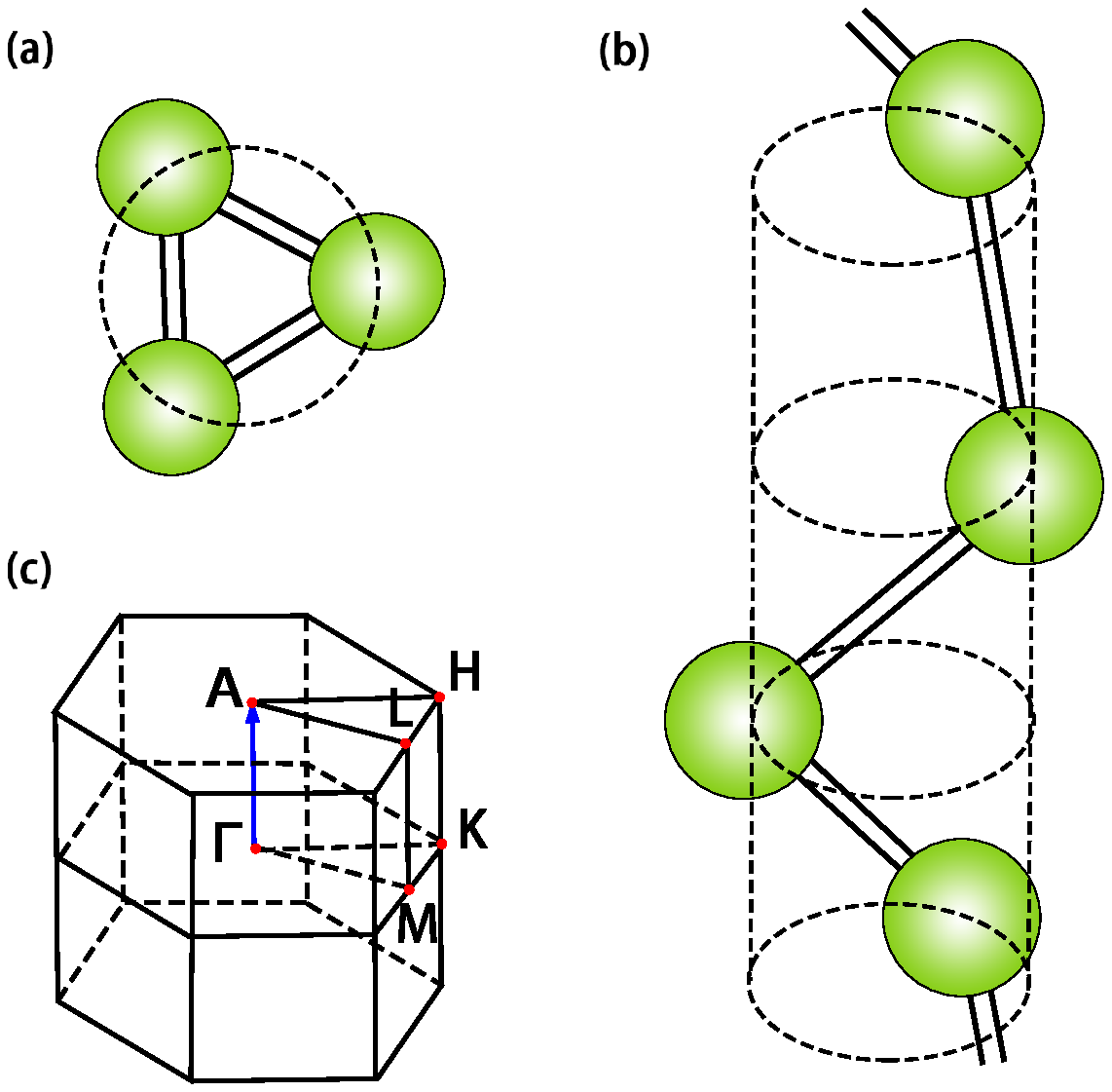}
\caption{\label{fig1}
(a) Top and (b) side views of one-dimensional selenium and tellurium
as well as (c) the associated $\vec{k}$ path in the calculation of band structure indicated by blue arrow.}
\end{figure}

\begin{figure}
\includegraphics[width=3in]{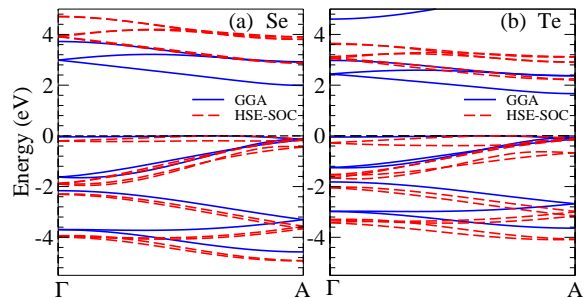}
\caption{\label{fig2}
Band structures of (a) selenium and (b) tellurium chains from the GGA (blue solid lines) calculations and HSE (red dashed lines) calculations with the SOC included.
Both materials possess indirect band gaps. The top of the valence bands is at 0 eV.}
\end{figure}

\begin{figure}
\includegraphics[width=3.33in]{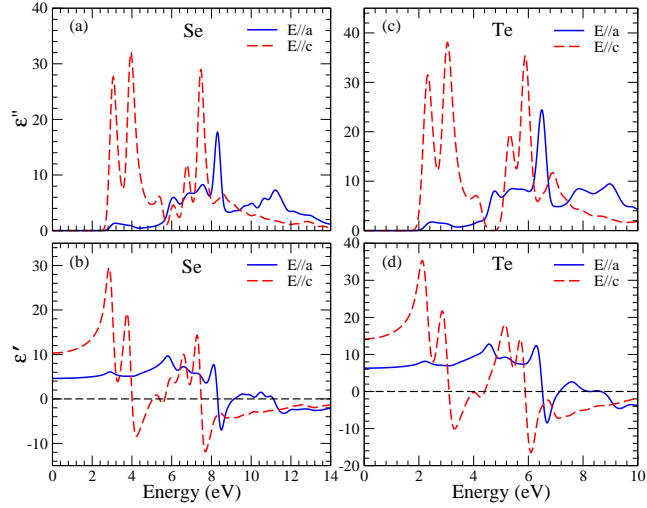}
\caption{\label{fig3}
Calculated imaginary [$\varepsilon''(\omega)$] and real [$\varepsilon'(\omega)$] part
of the dielectric function for (a) and (b) selenium chain as well as (c) and (d) tellurium chain for
light polarization perpendicular (E$\parallel$a) and parallel (E$\parallel$c) to the $c$ axis.}
\end{figure}

\begin{figure}
\includegraphics[width=3.33in]{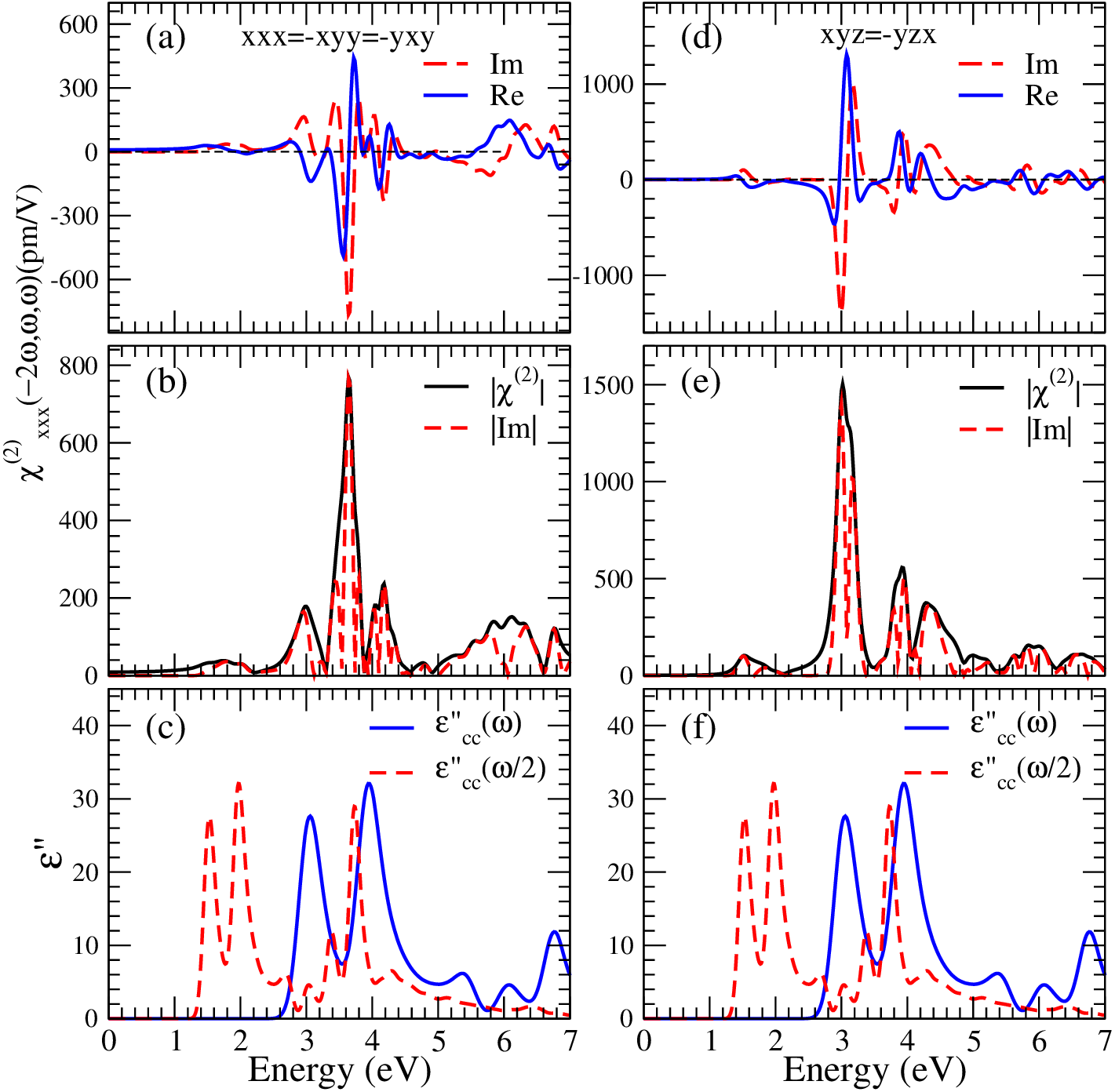}
\caption{\label{fig4}
(a) and (d) Real and imaginary parts of the second-order susceptibility for $\chi^{(2)}_{xxx}$ and $\chi^{(2)}_{xyz}$ in selenium chain, respectively. (b) and (e) Absolute value of the second-order susceptibility for $\chi^{(2)}_{xxx}$ and $\chi^{(2)}_{xyz}$ in selenium chain, respectively. (c) and (f) Imaginary part of the dielectric function for light polarization parallel to the $c$ axis.}
\end{figure}

\begin{figure}
\includegraphics[width=3.33in]{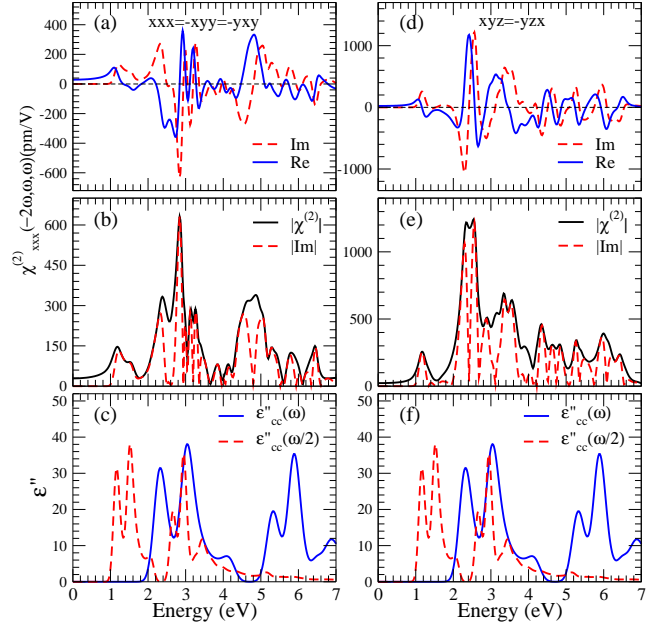}
\caption{\label{fig5}
(a) and (d) Real and imaginary parts of the second-order susceptibility for $\chi^{(2)}_{xxx}$ and $\chi^{(2)}_{xyz}$ in tellurium chain, respectively. (b) and (e) Absolute value of the second-order susceptibility for $\chi^{(2)}_{xxx}$ and $\chi^{(2)}_{xyz}$ in tellurium chain, respectively. (c) and (f) Imaginary part of the dielectric function for light polarization parallel to the $c$ axis.}
\end{figure}

\begin{figure}
\includegraphics[width=3in]{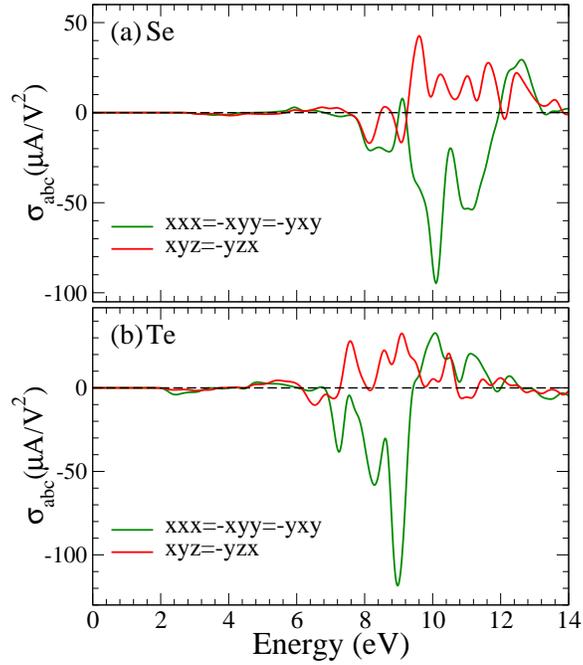}
\caption{\label{fig6}
The shift current response ($\sigma$) versus the photon energy for one-dimensional (a) selenium and (b) tellurium.}
\end{figure}

\begin{figure}
\includegraphics[width=3.33in]{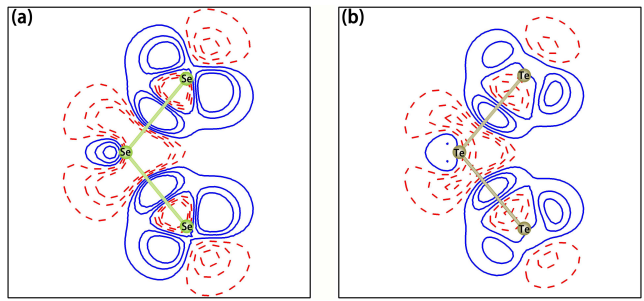}
\caption{\label{fig7}
The contour plots of the deformation charge densities for one-dimensional (a) selenium and (b) tellurium. The contour interval is 0.02 e/\AA$^3$.
The electron accumulation is depicted by positive contours (blue
solid lines), while the electron depletion is represented by negative contours (red dashed lines)}
\end{figure}
\vspace{5cm}


\newpage
\vspace{2cm}






\end{document}